\documentclass[letterpaper, 10 pt, conference]{ieeeconf} \IEEEoverridecommandlockouts                             

\usepackage{amssymb}
\usepackage{makecell}
\usepackage{amsmath}
\usepackage{color,xcolor}
\usepackage{algorithm, algpseudocode}

\usepackage[utf8]{inputenc}
\usepackage{booktabs, caption, makecell}

\usepackage{threeparttable}
\usepackage{bbm}
\usepackage{tikz}
\usetikzlibrary{positioning}
\usepackage{multirow}
\usepackage[figuresright]{rotating}
\usepackage{amsmath}

\usepackage{tablefootnote}

\usepackage{tikz-network}
\usepackage{algorithm, algpseudocode}

\usepackage{multirow}

\title{\LARGE \bf
A Latent Space Optimization Approach for Symbolic Discovery of Dynamical Models*
}

\author{Tongjia Liu$^{1}$, Ilias Mitrai$^{1,2}$% <-this % stops a space
\thanks{*This work was supported by the McKetta Department of Chemical Engineering.}% <-this % stops a space
\thanks{$^{1}$Tongjia Liu and Ilias Mitrai are with the McKetta Department of Chemical Engineering, The University of Texas at Austin, 78712, Austin, TX, USA}%%
\thanks{$^{2}$Corresponding author:  {\tt\small imitrai@che.utexas.edu}}}

\usepackage{lipsum}
\usepackage{cite}
\begin{document}

\maketitle
\thispagestyle{empty}
\pagestyle{empty}

\begin{abstract}

In this paper, we propose a hybrid latent-space Bayesian Optimization (BO) and global optimization framework for solving symbolic regression tasks to discover dynamical models from data. Symbolic regression can discover equations from data without fixing the functional form of the expression a priori. The proposed framework uses the fact that if the functional form of the expression is fixed, the symbolic regression task reduces to a parameter estimation problem, which must be solved to global optimality. Motivated by this structure, first, we train a VAE to map the discrete space of expression trees into a continuous latent space. Then, we use BO to search this latent space while assessing a candidate expression's prediction error by solving the parameter estimation problem to global optimality. We evaluate our framework across two case studies: static reaction rate law discovery and dynamic concentration identification in a continuous stirred-tank reactor (CSTR). The results show that our framework identifies the true governing equations faster than MINLP formulations without solver timeouts and achieves higher sample efficiency than evolutionary baselines under tight evaluation budgets.
\end{abstract}

\section{INTRODUCTION}
Symbolic regression is a data-driven technique that identifies mathematical expressions from data without making a priori assumptions on the functional form of the expression. By eliminating the need to assume a fixed functional form a priori, symbolic regression techniques simultaneously search the space of functional relationships between the input variables as well as the values of the associated parameters. Symbolic regression can yield models with high predictive accuracy while remaining interpretable, thereby providing physical and mechanistic insights into underlying process dynamics \cite{mckay1997steady,ly2012learning,engle2022deterministic}. Several computational approaches have been proposed to solve the symbolic regression task, i.e., search the discrete space of symbolic expressions and continuous space of parameters simultaneously. 

Traditionally, Genetic Programming (GP) and evolutionary algorithms have been used to search over candidate expression trees \cite{koza1994genetic, schmidt2009distilling}. Despite their widespread use, GP-based approaches perform a stochastic search and require many function evaluations. An alternative is to use neural network-based models, such as AI Feynman and graph neural networks, to simplify complex symbolic targets \cite{udrescu2020ai1,udrescu2020ai2,cranmer2020discovering}. AI Feynman trains a neural model to recursively identify functional symmetries, separability, and dimensional units to break down a complex problem into smaller subproblems. However, these methods require large datasets, assume that the unknown function is smooth, and often require domain-specific inputs like physical units. 

Alternatively, sparse regression techniques formulate equation discovery as a sparse feature selection problem \cite{brunton2016discovering}. These approaches rely on a predefined set of basis functions and seek a linear combination of the basis functions that minimizes the empirical prediction error \cite{cozad2014learning, narasingam2018data,mitrai2026learning}. Basis-function-based approaches, although computationally efficient, are inherently restricted by the predefined basis-function library; i.e., if the true governing terms are omitted from the candidate dictionary, they cannot be discovered. 

Recently, several mathematical programming approaches have been proposed to formulate the symbolic regression problem as a mixed-integer nonlinear optimization (MINLP) problem and solve it to global optimality with deterministic global optimization solvers \cite{cozad2018global,austel2017globally,kim2023learning, neumann2020new}. Despite providing deterministic global optimality guarantees, a primary limitation of existing MINLP formulations is scalability. The combinatorial search space grows exponentially with tree depth, the number of allowed operators, and the number of data points. Therefore, although systematic, applying global optimization techniques is computationally expensive. This challenge is exacerbated for the case of dynamical systems where the number of data points can be very large even if a single trajectory is available if the sampling time is low. An alternative to global optimization is using local optimization solvers. These solvers, although faster, can get trapped in local solutions and can be unreliable because their convergence depends strongly on the initial condition.

\begin{figure*}[htbp]
    \centering
    \includegraphics[trim = 0 105 0 0, clip, scale=0.45]{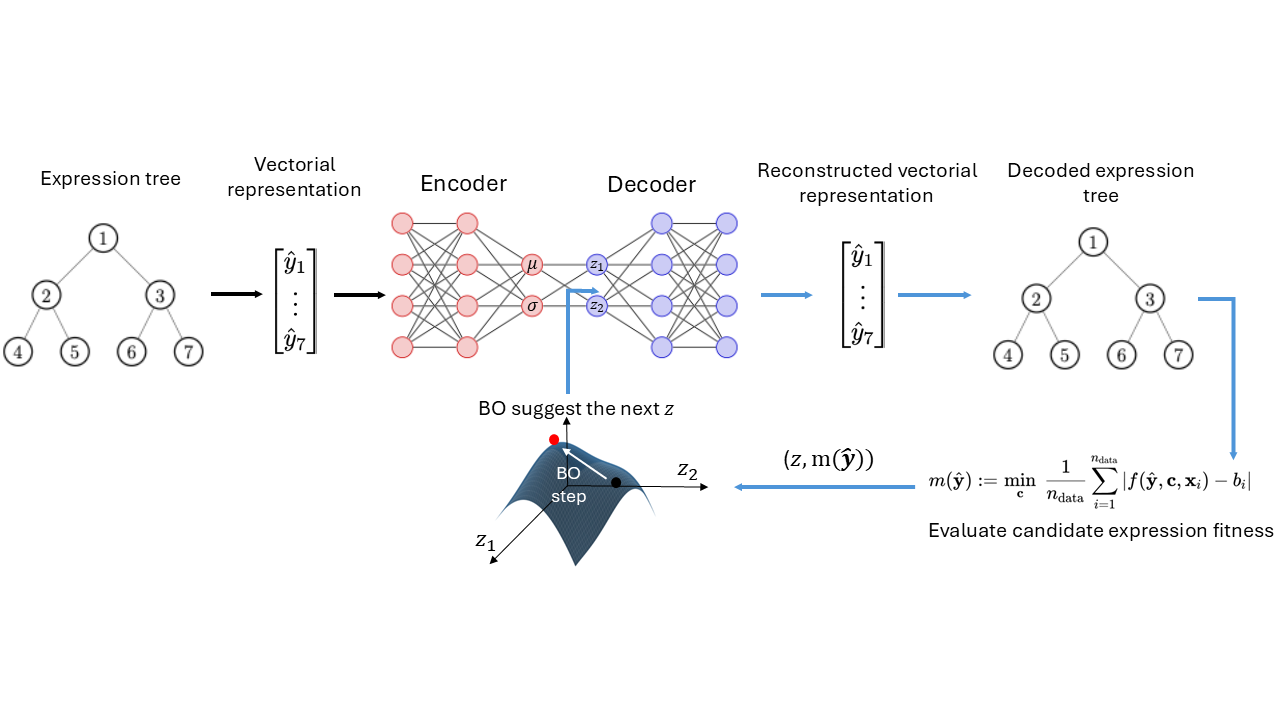}
    % \vspace{-60pt} % Adjust negative value to reduce space
    \caption{Proposed framework for symbolic regression. Blue arrows represent the Bayesian Optimization (BO) loop. }
    \label{fig:framework}
\end{figure*}

In this work, we propose a hybrid latent-space Bayesian Optimization (BO) and global optimization framework for obtaining high-quality primal solutions of the MINLP problem, i.e., functions with good predictive accuracy. As illustrated in Fig.~\ref{fig:framework}, the proposed framework is based on the observation that once the functional form is fixed, the symbolic regression task reduces to parameter estimation (i.e., fitting). However, the solution of parameter estimation problems to global optimality can be computationally expensive. We treat the symbolic regression task as a black box optimization problem and leverage Bayesian Optimization (BO) to navigate the space of functions while minimizing the number of function evaluations, i.e., the number of parameter estimation problems solved to global optimality.

The proposed approach has two components: first, a Variational Autoencoder (VAE) maps the discrete space of expression trees into a continuous latent space, transforming the combinatorial search into a continuous optimization task. Second, BO navigates this latent space to efficiently suggest candidate functions. We demonstrate our framework by discovering the governing reaction rate laws and dynamics of a continuous stirred-tank reactor (CSTR). The results show that our approach discovers the correct underlying equations while requiring less computational effort than existing mathematical programming methods and latent-space approaches that rely on genetic programming.

We note that recently, several latent-space approaches have been proposed to discover equations from data in symbolic form \cite{kusner2017grammar, tripp2020sample}. However, these approaches solve the parameter estimation problem, i.e., the function evaluation task, to local optimality and are used to discover static functions from physics, not dynamical models. The rest of this paper is organized as follows: In Section II, we present the MINLP formulation of the symbolic regression task, in Section III, we present the use of the VAE-BO approach for obtaining accurate expressions, i.e., high-quality primal solutions of the mixed-integer optimization problem, and in Section IV we present the computational results.

\section{Symbolic Discovery of Models via Mixed Integer Optimization} \label{sec: model}
Symbolic regression can be formulated as a deterministic MINLP by representing the candidate function as an expression tree \cite{cozad2018global,kim2023learning}. We assume that we are given a dataset $\mathcal{D} = \{\mathbf{x}_i, b_i\}_{i=1}^{n_{\text{data}}}$, generated by a function $f^{\text{true}}$, i.e., $b = f^{\text{true}}(\textbf{x})$, with $\mathbf{x}_i \in \mathbb{R}^{d_{\text{in}}}$ and $b_i \in \mathbb{R}$. The data can be generated by a static system, i.e., a steady-state model connecting $\textbf{x}$ and $b$ such as a reaction rate, or a dynamic one, i.e., a differential equation. The symbolic regression task seeks a $f \in \mathcal{F}$ that minimizes the Mean Absolute Error (MAE) as follows: $\min_{f \in \mathcal{F}}  \big(\sum_{i=1}^{n_{\text{data}}} |f(\mathbf{x}_i) - b_i|\big)/n_{\text{data}}.$

\subsection{Symbolic Representation of Target Expressions}
We represent $f$ as a binary expression tree of depth $d$ with $N = 2^{d+1}-1$ nodes. We define $\hat{\mathcal{N}} = \{1, \dots, N\}$ as the set of nodes and $\mathcal{T}=\{2^{d},\dots,2^{d+1}-1\}$ as the set of terminal (leaf) nodes. Each internal node $n \in \mathcal{N}_{\text{int}} = \hat{\mathcal{N}}\setminus \mathcal{T}$ has two children, nodes $2n$ and $2n+1$. At each node, a binary operator $\mathcal{B} = \{+, -, *, /\}$, a unary operator $\mathcal{U} = \{\sin, \cos, \exp, \log\}$, or an operand $\mathcal{L} = \{x_1, \dots, x_{d_{\text{in}}}, \text{cst}\}$ can be assigned, where $\text{cst}$ denotes continuous numerical constants. We denote $\mathcal{O} = \mathcal{B} \cup \mathcal{U} \cup \mathcal{L}$ as the set of all possible operators and operands that can be assigned to the nodes.

\subsection{Tree- and value-defining constraints}
We define the binary variable $y_{o,n} \in \{0, 1\}$ for each valid pair $(o, n) \in Y$, where $y_{o,n}$ is equal to one if $o$ is assigned to node $n$, and zero otherwise, with $Y := \{(o, n) \mid o \in O, n \notin T\} \cup \{(o, n) \mid o \in L, n \in T\}$. We also define $c_{n} \in [c^{\text{lb}}, c^{\text{ub}}]$ as the value of a constant assigned to node $n$. To ensure that the operator assignments yield a valid mathematical expression, it must satisfy four constraints \cite{cozad2018global,kim2023learning}. The first constraint guarantees that at most one operator or operand is assigned to each node, as shown below
\begin{equation}\label{eq:one_operator}
\sum_{o \in \mathcal{O}} y_{o,n} \le 1 \quad \forall n \in \mathcal{N}.
\end{equation}
The presence of an operator at a node dictates and constrains the assignment of operators to its children. For example, both unary and binary operators require a right child, whereas binary operators require a left child. These requirements are enforced via the following constraints
\begin{equation}\label{eq:operator_assignment_1}
\sum_{o \in \mathcal{B} \cup \mathcal{U}} y_{o,n} = \sum_{o \in \mathcal{O}} y_{o, 2n+1} \quad \forall n \notin \mathcal{T},
\end{equation}
\begin{equation}\label{eq:operator_assignment_2}
\sum_{o \in \mathcal{B}} y_{o,n} = \sum_{o \in \mathcal{O}} y_{o, 2n} \quad \forall n \notin \mathcal{T}.
\end{equation}
Finally, we enforce that at least one variable is present in the expression tree via the following constraint
\begin{equation}\label{eq:operator_assignment_3}
\sum_{n \in \mathcal{N}}\sum_{j=1}^{d_{in}} y_{x_j, n} \ge 1.
\end{equation}
This constraint renders expressions with only constants, e.g., $f(\textbf{x})= \text{cst}$, infeasible. 

The constraints presented so far guarantee that the expression tree is mathematically valid. The last set of constraints, called value-defining, computes the output of the expression tree for a given input and operator assignment. These constraints have the following general form and are written for all nodes $n$, data points $i$ and operators $o $
\begin{equation} \label{eq: value defining}
    G_{o}(v_{in}, v_{i,2n}, v_{i,2n+1}, y_{on}) \leq 0,
\end{equation}
where $v_{in}$ which is the value of node $n$ in the tree for data point $i$. For the case of multiplication, these constraints are equal to 
\begin{equation*}
\begin{aligned}
    v_{in} - v_{i,2n}v_{i,2n+1} & \leq \overline{M}(1-y_{*,n})\\
    v_{in} - v_{i,2n}v_{i,2n+1} & \geq \underline{M}(1-y_{*,n}).
\end{aligned}
\end{equation*}

Overall, the resulting symbolic regression task is
\begin{equation}\label{eq:symbolic_regression_minlp}
\begin{aligned}
    \min_{y_{on}, c_{n}, v_{in}} \ & \frac{1}{n_{\text{data}}} \sum_{i=1}^{n_{\text{data}}} |f(y, c, \mathbf{x}_i) - b_i| \\
    \text{s.t.} \ & \text{Eqs. } \eqref{eq:one_operator}, \eqref{eq:operator_assignment_1}, \eqref{eq:operator_assignment_2}, \eqref{eq: value defining}.
\end{aligned}
\end{equation}
This is a mixed-integer nonlinear programming problem due to the value-defining constraints, which are nonconvex for most operators, e.g., multiplication and division. 

\section{Black-Box Optimization for Symbolic Regression}
The computational complexity of the symbolic regression MINLP problem depends on the tree depth, the number and type of operators, and the number of data points. We observe that if the binary variables are fixed, then the resulting symbolic regression problem is a continuous nonlinear parameter estimation problem, i.e., the only degrees of freedom are the continuous variables $c_{n}$. Motivated by this structure, we adopt a data-driven approach to navigate the discrete space of symbolic expressions. First, we use a variational autoencoder to learn a continuous parameterization of the discrete symbolic space of differential equations. 
\subsection{Continuous Representation via Variational Autoencoders}

To transform the discrete space of symbolic expressions into a continuous one, we first convert the binary variables $\{y_{o,n}\}$ into an integer vector. Then, we order the set of operators $o \in \mathcal{O}$ and convert $y_{o,n}$ into integer variables $\hat{y}_{n} \in \{1,\dots, |\mathcal{O}|\}$ for each node $n \in \mathcal{N}$, defining the vector $\hat{\mathbf{y}} = \{\hat{y}_{n}\}_{n \in \mathcal{N}}$. The VAE architecture consists of an encoder and a decoder. The encoder maps the integer representation $\hat{\mathbf{y}}$ onto the parameters of a multivariate Gaussian distribution
\begin{equation}
    (\boldsymbol{\mu}, \boldsymbol{\sigma}^2) = \mathrm{Enc}(\hat{\mathbf{y}}; \theta_{e}),
\end{equation}
where $\boldsymbol{\mu}, \boldsymbol{\sigma}^2 \in \mathbb{R}^{N_{h}}$, with $N_{h}$ being the dimension of the latent space, and $\theta_{e}$ denoting the encoder's learnable parameters. A latent vector $\mathbf{z} \in \mathbb{R}^{N_h}$ is sampled via the reparameterization trick $\mathbf{z} = \boldsymbol{\mu} + \boldsymbol{\sigma} \odot \boldsymbol{\varepsilon}$, with $\boldsymbol{\varepsilon} \sim \mathcal{N}(\mathbf{0}, \mathbf{I})$. 
The decoder network performs the inverse mapping to reconstruct the expression tree:
\begin{equation}
    \hat{\mathbf{y}}^{\mathrm{pred}} = \mathrm{Dec}(\mathbf{z}; \theta_{d}),
\end{equation}
where $\theta_{d}$ denotes the learnable parameters of the decoder.

To ensure unbiased representation across arbitrary functional forms, the VAE is trained on a dataset of random and syntactically valid expression trees, i.e., random trees that satisfy arity constraints in Eqs.~\eqref{eq:one_operator}--\eqref{eq:operator_assignment_3}. The VAE weights are trained by minimizing the following loss function:
\begin{equation}
    \mathcal{L}_{\text{total}} = \mathcal{L}_{\text{recon}} + \beta \mathcal{L}_{\text{KL}},
\end{equation}
where $\mathcal{L}_{\text{recon}}$ represents the cross-entropy reconstruction loss between $\hat{\mathbf{y}}$ and $\hat{\mathbf{y}}^{\mathrm{pred}}$, $\mathcal{L}_{\text{KL}}$ computes the Kullback-Leibler divergence relative to a standard normal prior $\mathcal{N}(\mathbf{0}, \mathbf{I})$, and $\beta$ is a scaling hyperparameter controlling latent space smoothness.

\subsection{Evaluating the Fitness of an Expression Tree}
For a fixed functional form parameterized by $\hat{\mathbf{y}}$, i.e., 
expression tree structure, the computation of the empirical prediction error requires the fitting of the constants $c_{n}$. This corresponds to the following continuous nonlinear optimization problem
\begin{equation}\label{eq:expression_evaluation}
\begin{aligned}
    m(\hat{\mathbf{y}}) := \min_{\mathbf{c}} \ & \frac{1}{n_{\text{data}}} \sum_{i=1}^{n_{\text{data}}} |f(\hat{\mathbf{y}}, \mathbf{c}, \mathbf{x}_i) - b_i| \\
    \text{s.t.} \ & c^{\text{lb}} \le c_{n} \le c^{\text{ub}} \quad \forall n \in \mathcal{N},
\end{aligned}
\end{equation}
where $m(\hat{\textbf{y}})$ denotes the prediction error of the symbolic expression fitted in the training data. We solve this problem to global optimality. If a decoded expression leads to domain violations (e.g., division by zero or logarithm of non-positive values), the evaluation is considered infeasible, and a large constant penalty is assigned to $m(\hat{\mathbf{y}})$.

\subsection{Latent-Space Bayesian Optimization}
Given the learned continuous latent space and the evaluation problem in Eq.~\ref{eq:expression_evaluation}, we use BO to find accurate expressions. In this approach, a Gaussian Process (GP) with a Matérn kernel is used to approximate $\mathcal{F}(\textbf{z}) = m(\hat{\mathbf{y}}(\textbf{z}))=m(\hat{\mathbf{y}})$ and Log Expected Improvement (LogEI) as the acquisition function.

The optimization workflow proceeds by first initializing the GP surrogate using a dataset of randomly sampled valid trees $\{\hat{\mathbf{y}}_i\}$, their latent space representation $\{\mathbf{z}_i\}$, and their evaluation $\{m(\hat{\mathbf{y}}_i)\}= \{\mathcal{F}(\textbf{z}_{i})\}$. At each BO iteration, the acquisition function is maximized to identify a candidate coordinate $\mathbf{z}^* \in \mathcal{Z}$, which is then passed through the decoder network using an $\mathrm{argmax}$ operation to yield the discrete expression tree $\hat{\mathbf{y}} = \mathrm{Dec}(\mathbf{z}^*; \theta_d)$. Next, we check whether $\hat{\mathbf{y}}$ satisfies the arity constraints in Eqs.~\eqref{eq:one_operator}--\eqref{eq:operator_assignment_2}. If the candidate tree is invalid, continuous parameter $\mathrm{cst}$ optimization is bypassed and a penalty score is assigned to $m(\hat{\mathbf{y}})$; otherwise, the continuous NLP in Eq.~\eqref{eq:expression_evaluation} is solved to evaluate $m(\hat{\mathbf{y}})$. Finally, the GP surrogate is updated with the evaluation pair $(\mathbf{z}^*, m(\hat{\mathbf{y}}))$, and the process repeats until a predefined termination criterion is met.

\section{COMPUTATIONAL RESULTS}\label{sec res}
To evaluate search efficiency, structural recovery, and noise robustness of the proposed approach, we consider two case studies: 1) {Case Study A:} Discovery of a nonlinear reaction rate law from concentration data; 2) {Case Study B:} Discovery of governing equations for a CSTR with varying flowrate and a first-order reaction.

\subsection{VAE Training and Latent Dimension Sensitivity}
In both case studies, the number of input variables is two, and we use the same tree depth and number of operators. Therefore, we train one VAE. Specifically, we fix the tree depth to three ($d=3$), which leads to expression trees with $15$ nodes. The set of allowable binary operators is $\{+, -, *\}$ and operands $\{x_{1}, x_{2}, \mathrm{cst}\}$. Each expression tree is encoded as an integer vector $\hat{\mathbf{y}} = \{\hat{y}_n\}_{n \in \mathcal{N}} \in \{1, \dots, \vert{}\mathcal{O}\vert{}\}^{15}$. The vector $\hat{\mathbf{y}}$ is the input to a 10-dimensional embedding layer which yields a flattened 150-dimensional encoder input vector. The encoder processes this input through a 256-unit hidden layer with ReLU activation to project mean $\boldsymbol{\mu}$ and log-variance $\log \boldsymbol{\sigma}^2$ vectors. The decoder mirrors the encoder structure with a 256-unit ReLU hidden layer, projecting back through a linear output layer to recover node-wise categorical probability distributions. The network was trained for 100 epochs using the Adam optimizer \cite{kingma2014adam} with a learning rate of $10^{-3}$, a batch size of 512, and $\beta=0.05$.

\begin{figure}[htbp]
    \centering
    \includegraphics[width=0.9\linewidth]{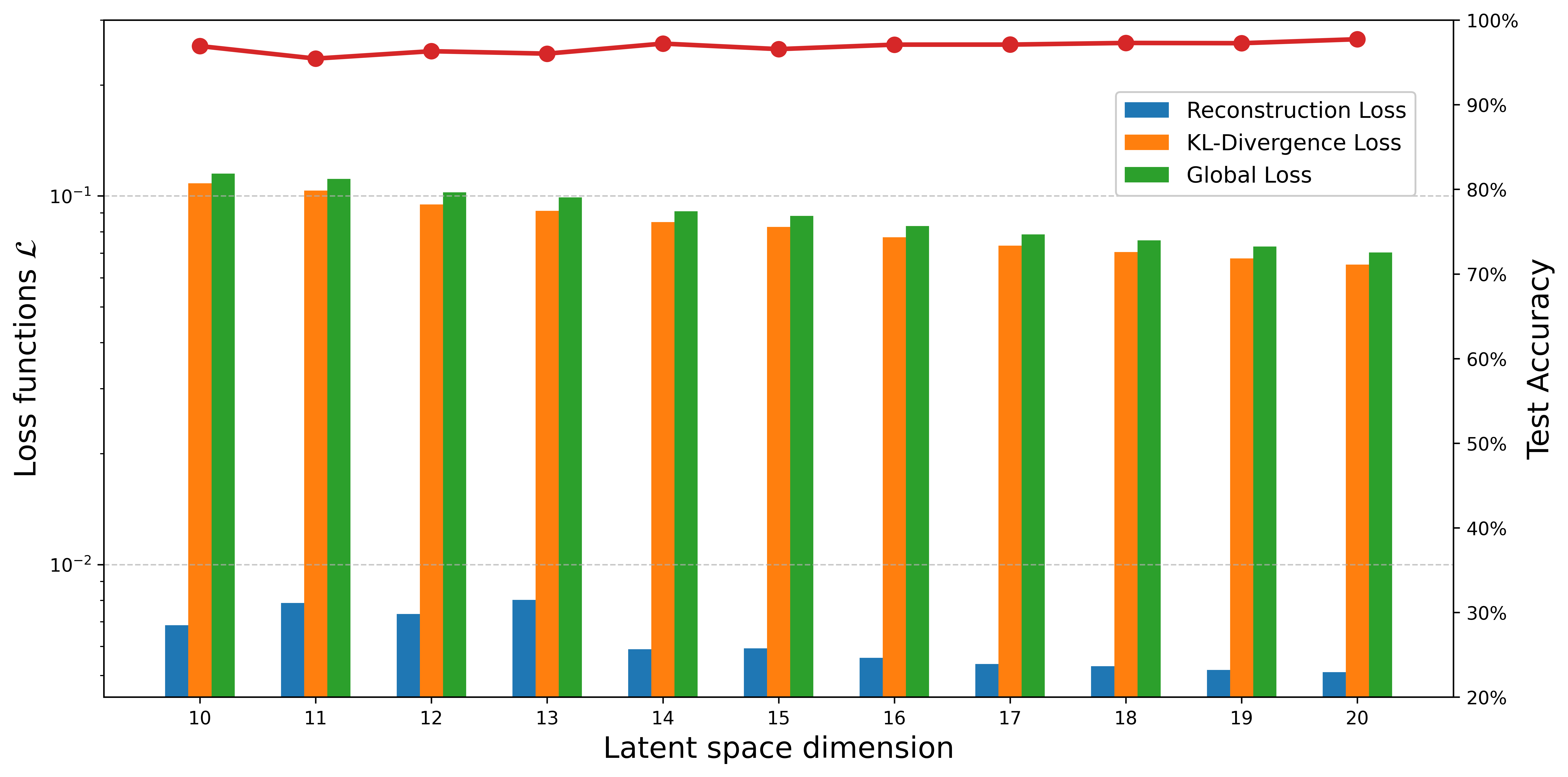}
    \caption{VAE training performance and tree reconstruction accuracy on the test set (red line) across varying latent dimensions $d_z$.}
    \label{fig:latent_sensitivity}
\end{figure}

To identify an appropriate latent dimension $d_z$, we performed a sensitivity analysis across $d_z \in \{10, 20\}$. Fig.~\ref{fig:latent_sensitivity} illustrates the trade-off between loss components (Reconstruction, KL-Divergence, and Global Loss) and tree reconstruction accuracy on test data.
From Fig.~\ref{fig:latent_sensitivity}, increasing $d_z$ from 10 to 20 reduces both KL-divergence and global loss. Reconstruction loss drops greatly between $d_z = 10$ and $d_z = 14$, plateauing near $4 \times 10^{-3}$. Reconstruction accuracy (red line in Fig.~\ref{fig:latent_sensitivity}) remains above 95\% across all candidate dimensions, peaking at $d_z = 15$ ($\approx 97.5\%$). Selecting $d_z > 15$ yields minimal gains in reconstruction accuracy. Therefore, $d_z = 15$ was selected to balance expression reconstruction accuracy with BO search efficiency.

\subsection{Case Study A: Static Reaction Rate Law Discovery}

\subsubsection{Problem Definition and Data Generation}
In this case study, we want to discover the governing rate law for a reversible reaction ($2\text{A} \rightleftharpoons \text{B}$) using concentration-rate observations. State observations were sampled across the concentration space of reactants A ($C_{\text{A}} \equiv x_1$) and B ($C_{\text{B}} \equiv x_2$). Specifically, we generated $N_{\text{data}} = 200$ data uniformly over $C_{\text{A}}, C_{\text{B}} \sim \mathcal{U}(0.1, 2.5)\text{M}$. The true net reaction rate $r_{\text{net}}$ follows second-order forward and first-order reverse kinetics:
\begin{equation}
r_{\text{net}}(C_{\text{A}}, C_{\text{B}}) = k_1 C_{\text{A}}^2 - k_2 C_{\text{B}}
\end{equation}
where $k_1 = 0.50$ and $k_2 = 0.20$ represent the ground-truth kinetic rate constants. Zero-mean Gaussian noise ($\sigma = 0.01$) was added to simulate experimental uncertainty, i.e., $r_{\text{noisy}} = r_{\text{net}} + \epsilon, \quad \epsilon \sim \mathcal{N}(0, 0.01^2)$. The dataset comprises feature matrix $\mathbf{X} \in \mathbb{R}^{200 \times 2}$ and noisy target vector $\mathbf{r} \in \mathbb{R}^{200 \times 1}$.

\subsubsection{Proposed Framework Performance}
We initialized the search with 30 randomly sampled expression trees to construct the initial GP surrogate model and then initiated BO. During fitness evaluation of a candidate tree structure, constant node bounds were set to $c_{\text{lo}} = -50$ and $c_{\text{up}} = 50$. We allowed 50 total function evaluations, which required 82.23 seconds: 35.0 seconds for parameter fitting (evaluation) using BARON and 47.2 seconds for GP surrogate updates and acquisition function optimization. The lowest-loss expression structure was discovered on the first BO iteration (candidate evaluation \#31 overall). The best expression identified achieved an objective loss of $0.00781$ and was equal to
\begin{equation}
r_{\text{pred}} = 0.4988 C_{\text{A}}^2 - 0.2012 C_{\text{B}} + 0.0033 C_{\text{A}}
\end{equation}
The proposed approach correctly recovered the true kinetic functional form ($0.50 C_{\text{A}}^2 - 0.20 C_{\text{B}}$). However, a linear term with a small constant $(0.0033)$ is present due to the noise. 

\begin{figure}[htbp]
    \centering
    \includegraphics[width=0.8\linewidth]{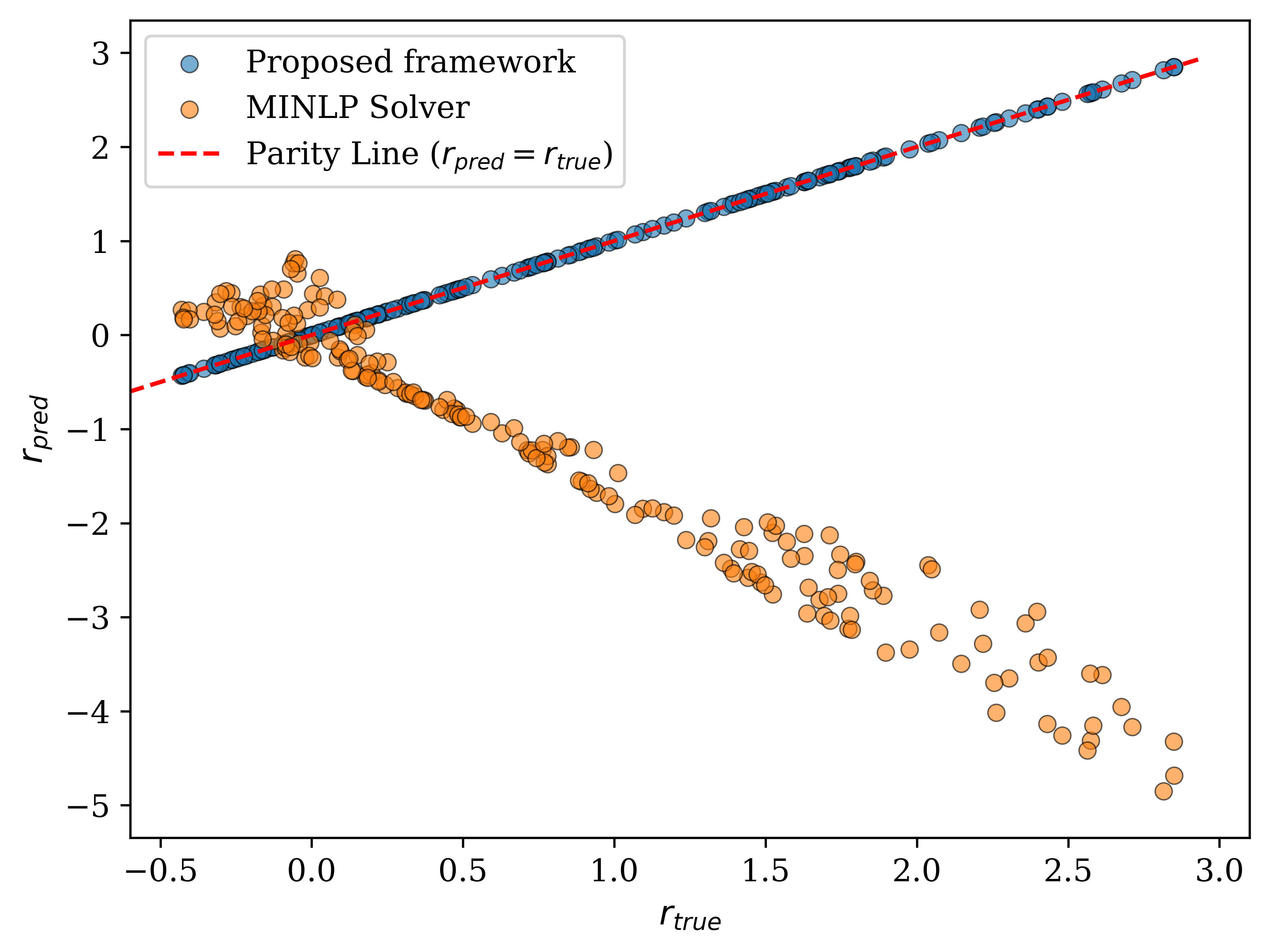}
    \caption{Comparison of predicted reaction rate versus test data for case study A.}
    \label{fig:case1_results}
\end{figure}

\subsubsection{Comparison with baselines}
We compare the latent space BO against solving the deterministic MINLP formulation from \cite{kim2023learning} with BARON v.25.8.5 \cite{zhang2025solving} using the same tree depth, number of data points, and operators as described above. The MINLP model has 41,045 constraints and 6,482 variables (including 66 binary variables). Given a CPU time limit of 3,600 seconds, BARON completed only 16 branch-and-reduce iterations before timing out. The best expression found has an objective value equal to $0.7790$. In comparison, the proposed approach achieved a substantially lower objective value ($0.00781$), in significantly less computing time (82.23~s vs. 3,600~s). The predictions obtained with the function discovered by the MINLP formulation are plotted in Figure~\ref{fig:case1_results} alongside the ground-truth observations and the predictions obtained from the proposed approach. 

\subsection{Case Study B: Dynamic Isothermal CSTR System Identification}

\subsubsection{Problem Definition and Data Generation}
In the second case study, we consider the discovery of the mass balance for an isothermal CSTR in which an irreversible reaction $\text{A} \rightarrow \text{B}$ occurs. The dynamic behavior of the system is described by the following differential equation
\begin{equation}\label{eq:cstr_mass_balance}
\frac{\mathrm{d}C(t)}{\mathrm{d}t} = \frac{F(t)}{V}(C_{\text{f}} - C(t)) - kC(t)
\end{equation}
where $C(t)\equiv x_{1}$ is the reactor concentration, $F(t) \equiv x_{2}$  is the volumetric feed flow rate, $V = 1\text{ L}$ is the volume, $C_{\text{f}} = 2\text{ M}$ is the feed concentration, and $k = 2\text{ s}^{-1}$ is the reaction constant. Substituting these values into Eq.~\ref{eq:cstr_mass_balance} yields the following differential equation (using the $x_{1}$ and $x_{2}$ notation):
\begin{equation}\label{eq:cstr_analytical}
\dot{x}_1 = 2.000 x_2 - x_2 x_1 - 2.000 x_1.
\end{equation}
We generate training and testing data by simulating the system from five initial concentrations $C_0 \in \{0.1, 0.5, 1.0, 2.0, 3.5\}\text{ M}$ over $t \in [0, 2.0]\text{ s}$ with $\Delta t = 0.01$ ($201$ time points per trajectory and $N = 1,005$ total observations) using a RK45 explicit Runge-Kutta. The feed flowrate is equal to $F(t) = 6 + \sin(10t)$. Gaussian noise ($\sigma = 0.01$) was added to $C(t)$, and the time derivatives $\frac{\mathrm{d}x_1}{\mathrm{d}t}$ were estimated using second-order finite differences via \texttt{numpy.gradient}. The dataset consists of state matrices $\mathbf{X}^{(i)} \in \mathbb{R}^{1005 \times 2}$ and target derivative vectors $\dot{\mathbf{x}}_1^{(i)} \in \mathbb{R}^{1005 \times 1}$.

\subsubsection{Proposed Framework Performance}
We initialize the proposed framework with 30 random expressions. After 40 iterations ($30 + 40 = 70$ total evaluations) the following differential equation was found
\begin{equation}\label{eq:proposed_result}
\dot{x}_{1,\text{proposed}} = 2.010 x_2 - 2.045 x_1 - x_2 x_1 + 0.000491.
\end{equation}
Comparing Eq.~\eqref{eq:proposed_result} to the mass balance in Eq.~\eqref{eq:cstr_analytical}, the proposed approach correctly identified the bilinear state-input coupling ($-x_2 x_1$) and closely estimated the physical parameters ($2.010 \approx 2.0$ feed coefficient, $-2.045 \approx -2.0$ decay rate). The residual objective loss of $0.000491$ is caused by noise in the derivative data. We note that BO found the above expression after 20 iterations (50 total evaluations). Overall, the computational time, including BO iterations and initialization, was 164.76 seconds, 119.73 seconds required for optimizing the constants via BARON and 44.37 seconds for updating the GP and optimizing the acquisition function.

\subsubsection{Comparison with baselines}
We compare the proposed latent-space BO approach with EDHiE \cite{mevznar2023efficient}, which uses hierarchical VAEs to learn a continuous representation, searches the continuous space using evolutionary algorithms, and fits the constants for each expression in the population using local optimization methods. To ensure fair comparison regarding the total number of function evaluations, we provide both approaches with 100 total function evaluations and run them across 10 different seeds (seeds 18–27) as follows:
\begin{enumerate}
    \item \textbf{Proposed Framework (BO+VAE):} Generate 30 initial random samples followed by 70 Bayesian optimization iterations ($30 + 70 = 100$ maximum tree evaluations per seed).
    \item \textbf{EDHiE (HVAE + EA Baseline):} Configured with a population size of 10 over 10 generations ($10 \times 10 = 100$ candidate evaluations per seed) and a success threshold of $10^{-5}$.
\end{enumerate}
Across the 10 runs using different random seeds, EDHiE always failed to recover the exact ground-truth dynamic expression. The best expression discovered by EDHiE, across all seeds, is 
\begin{equation}\label{eq:eidhe_result}
\dot{x}_{1,\text{EDHiE}} = 1.949 x_2 - 1.292 x_2 x_1.
\end{equation}
When run without a budget limit, EDHiE required about 1,500 candidate function evaluations to discover the ground-truth expression.

\begin{figure}[t]
    \centering
    \includegraphics[width=0.9\linewidth]{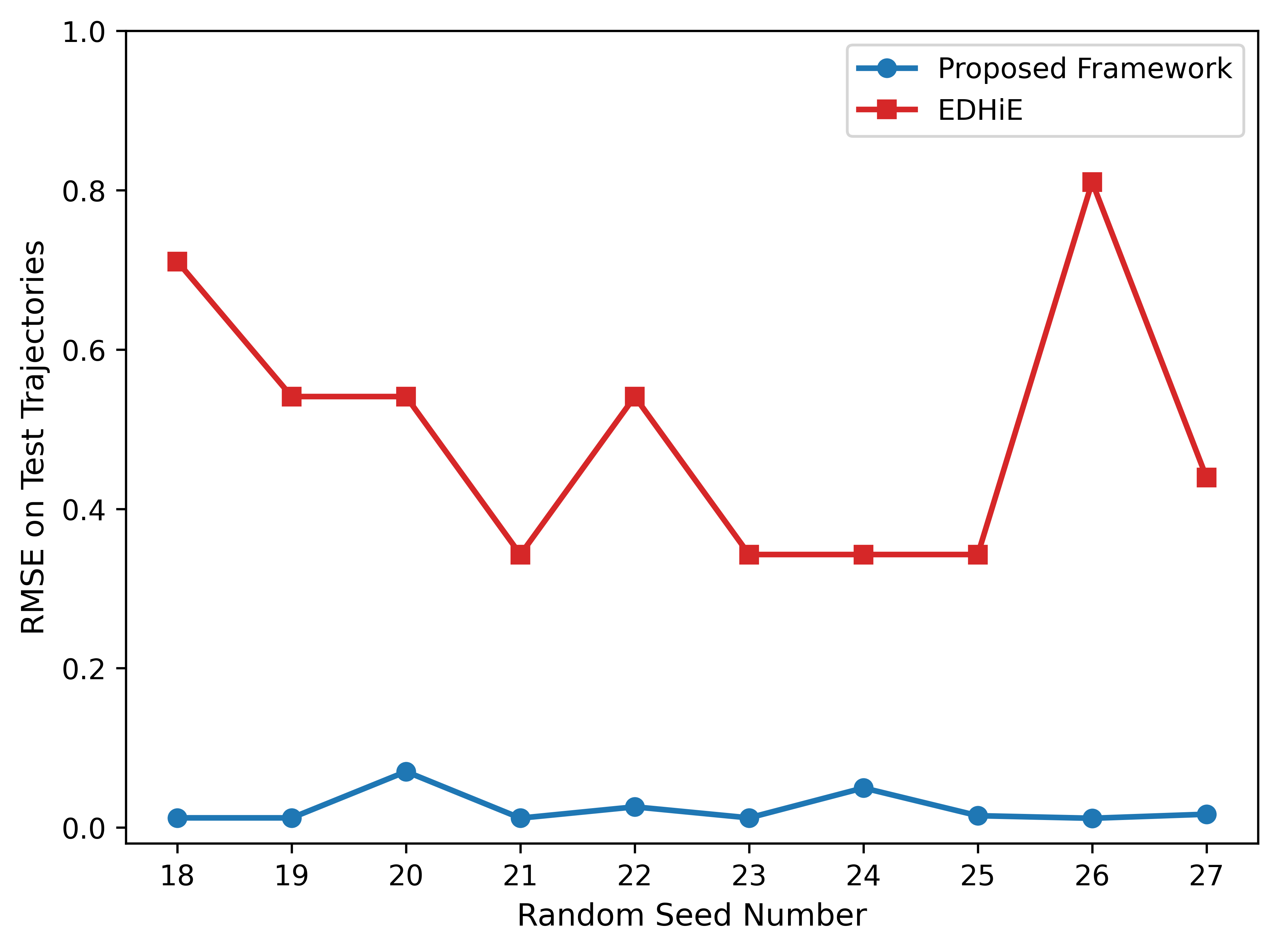}
    \caption{RMSE on test set across 10 independent random seeds with 100 evaluations per run for the proposed approach and EDHiE.}
    \label{fig:case2_multi_seed}
\end{figure}

For a direct comparison, we evaluated the proposed framework across the same 10 random seeds. The proposed framework successfully recovered the correct ground-truth dynamic expression in 7 out of 10 runs. In the 3 runs that did not achieve exact structural recovery, the proposed approach still achieved a lower RMSE on the test set than EDHiE, as shown in Figure~\ref{fig:case2_multi_seed}.

Overall, these results demonstrate that the proposed framework is more sample-efficient than HVAE paired with evolutionary algorithms. Evolutionary algorithms rely on evaluating large populations over many generations, assuming parameter optimization is cheap. However, solving the parameter estimation problem to global optimality can be computationally expensive. In contrast, the proposed framework uses BO to navigate the continuous latent space and obtain high-quality results even with a few evaluations.

Finally, we evaluate the predictive accuracy of the model obtained with the proposed approach and EDHiE using an out-of-distribution profile for the inlet flowrate, which was set equal to $F_{\text{test}}(t) = 5.25 + 1.5\sin(4t) + 0.8\cos(12t)$, and the initial concentration $C(0) = 0.2\text{M}$.

\begin{figure}[t]
    \centering
    \includegraphics[width=0.9\linewidth]{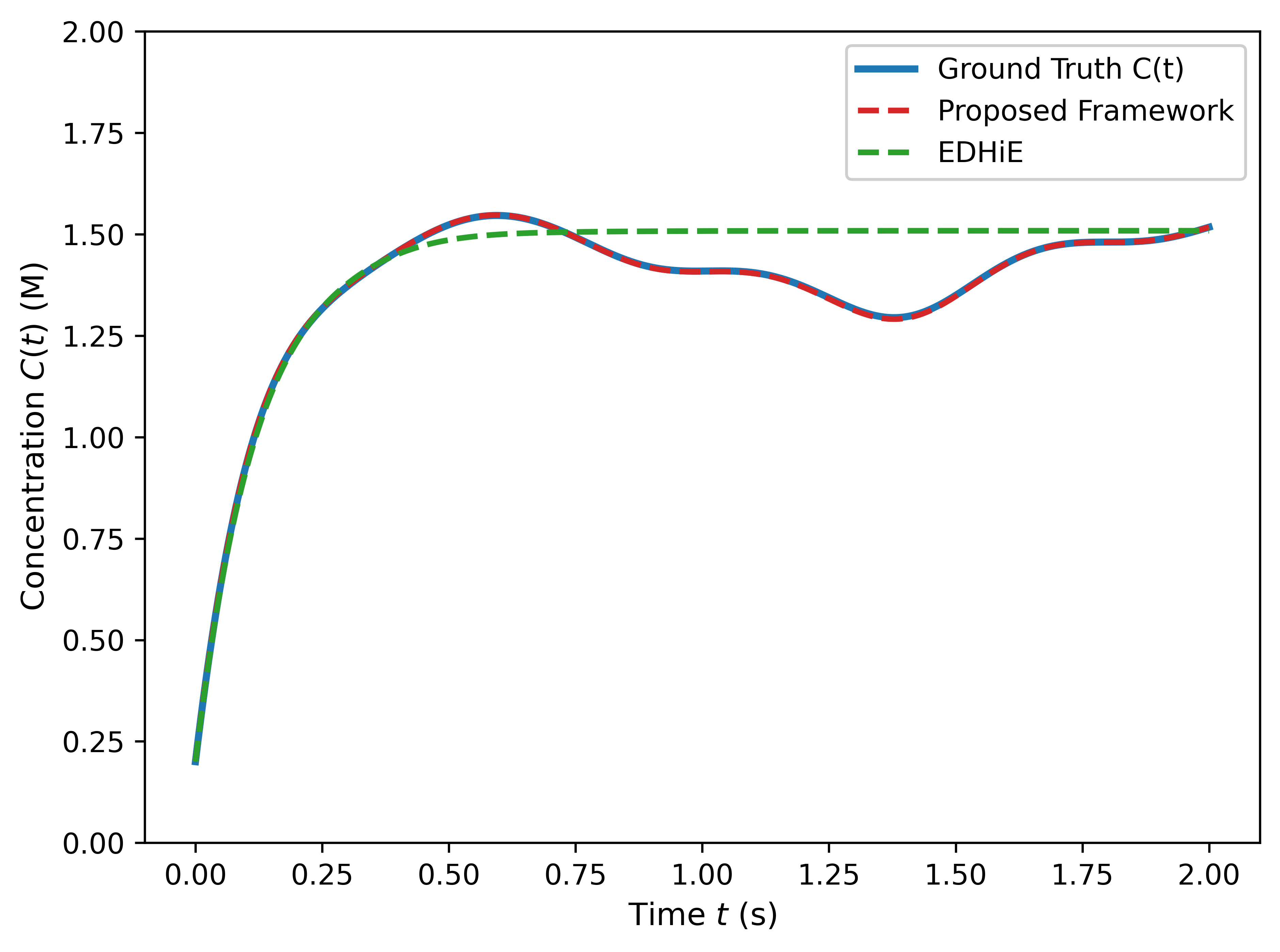}
    \caption{Out-of-distribution concentration $C(t)$ obtained by RK4 integration of the differential equations discovered by each model under unseen flow rate $F_{\text{test}}(t)$}
    \label{fig:case2_trajectories}
\end{figure}

The prediction of the reactant concentration with time for the model obtained with the proposed approach and EDHiE is presented in Fig.~\ref{fig:case2_trajectories}. The results show that the temporal profile of the reactant concentration obtained with the proposed approach matches the profile from the true model across the entire simulation horizon. In contrast, predictions from the EDHiE-derived model match the initial growth only up to $t \approx 0.5\text{ s}$ before stalling into a flat plateau at $C \approx 1.5\text{ M}$, since the identified expression lacks the self-stabilizing $-2x_1$ decay term. In summary, these results demonstrate that continuous learned representations of expression trees combined with Bayesian optimization provide better sample efficiency over standard evolutionary algorithms, enabling the reliable recovery of governing equations, especially for cases where the parameter estimation problem is solved to global optimality.

\section{Conclusions}
In this work, we present a latent-space Bayesian Optimization framework to solve symbolic regression problems. We represent candidate functions as expression trees to capture discrete equation structures. To efficiently navigate this space, first, we learn a continuous representation of expression trees using a Variational Autoencoder, which transforms the discrete equation discovery into a continuous optimization task. This continuous representation is then searched using Bayesian Optimization. We evaluated the proposed framework across two case studies. In the first case study, the framework discovered a reaction rate law from observational data. In the second case study, it identified the governing equation for an isothermal CSTR. Compared to deterministic mathematical programming approaches, our framework identified the governing equations in significantly less computation time. Additionally, our proposed framework is more sample-efficient than evolutionary algorithm baselines, consistently uncovering the ground truth function within a tight evaluation budget.

\addtolength{\textheight}{-12cm}   % This command serves to balance the column lengths

\vspace{10pt}
\addcontentsline{toc}{section}{References}
\bibliographystyle{ieeetr}
\bibliography{bib_sample}

\end{document}